\documentclass[aps,pra,twocolumn,groupaddress,floatfix]{revtex4-1}
\usepackage[svgnames]{xcolor}
\usepackage[colorlinks=true,urlcolor = violet,linkcolor=blue,citecolor=teal,bookmarks=false]{hyperref}
\usepackage{amsfonts,amsmath,amssymb}
\usepackage{graphicx}
\usepackage{dsfont}

\usepackage{tikz}
\usetikzlibrary{arrows,topaths,shapes.geometric,decorations.markings,plotmarks,shadows,positioning,plotmarks,matrix}
 
\usepackage{pgfplots}

\def\ii{{\rm i}}
\newcommand{\dd}{{\rm d}}
\newcommand{\DD}{\mathfrak{D}}

\def\expect#1{\langle#1\rangle}

\begin{document}

\title{Domain wall dynamics in the Landau--Lifshitz magnet\\
and the classical-quantum correspondence for spin transport}

\author{Oleksandr Gamayun}
\affiliation{Institute for Theoretical Physics and Delta Institute for Theoretical Physics,
University of Amsterdam, Science Park 904, 1098 XH Amsterdam, The Netherlands}

\author{Yuan Miao}
\affiliation{Institute for Theoretical Physics and Delta Institute for Theoretical Physics,
University of Amsterdam, Science Park 904, 1098 XH Amsterdam, The Netherlands}

\author{Enej Ilievski}
\affiliation{Institute for Theoretical Physics and Delta Institute for Theoretical Physics,
University of Amsterdam, Science Park 904, 1098 XH Amsterdam, The Netherlands}

\date{\today}

\begin{abstract}
We investigate the dynamics of spin in the axially anisotropic Landau--Lifshitz field theory with a magnetic domain wall initial 
condition. Employing the analytic scattering technique, we obtain the exact scattering data and reconstruct the time-evolved
profile. We identify three qualitatively distinct regimes of spin transport,
ranging from ballistic expansion in the easy-plane regime, absence of transport in the easy-axis regime
and logarithmically enhanced diffusion for the isotropic interaction. Our results are in perfect qualitative agreement with
those found in the anisotropic quantum Heisenberg spin-$1/2$ chain, indicating a remarkable classical-quantum 
correspondence for macroscopic spin transport.
\end{abstract}


\maketitle


\paragraph*{Introduction.}

The theory of exactly solvable partial differential equations~\cite{Faddeev_book,Ablowitz_book,Novikov_book,Drazin_book}, colloquially
known as the theory of solitons~\cite{ZK65}, represents one of the cornerstones of theoretical and mathematical physics.
While the technique has been traditionally used mostly as a theoretical framework to describe various nonlinear wave phenomena such as 
dispersive shock waves~\cite{GP74,El16} and modulational instabilities~\cite{Hasegawa84,Tracy84,Ercolani90},
soliton systems also played an instrumental role in a broader range of physics applications, ranging
from experimentally relevant setups with cold atoms and BECs~\cite{Kevrekidis_book}, ocean waves~\cite{Osborne02},
physics of plasmas and nonlinear media~\cite{PhysRevLett.102.135002}, Josephson junctions
and nonlinear optics~\cite{Hasegawa_book,Malomed_book,optic}, and many theoretical concepts including the AdS/CFT 
correspondence~\cite{MZ03,KMMZ04}, Gromov-Witten theory~\cite{Marshakov08}, 
Painlev\'e transcendents~\cite{Fokas_2006,Clarkson_2003,Gamayun_2013} and random matrix theory~\cite{TracyWidom94,TracyWidom11}.

Exact results on the nonequilibrium properties of soliton systems, both near and far from equilibrium, are nonetheless extremely rare.
This can attributed to the fact that, outside of a few exceptional cases~\cite{GBC15,Gamayun_2016S,CD16,De_Luca_2016}, the formal integration scheme cannot 
be implemented in a fully analytic manner in general. For this reason, in physics application one mostly relies on linearization or 
various approximations \cite{RevModPhys.61.763} and asymptotic techniques~\cite{AS77,DZ93,DIZ93}.
In this Letter, we identify an exceptional but physically relevant nonequilibrium scenario where the issue can be overcome.
We consider the Landau--Lifshitz ferromagnet and calculate the exact nonlinear Fourier spectrum (scattering data)
for the magnetic domain wall initial profile. This enables us to analytically explore its far-from-equilibrium transport properties.
We study the time-evolution of the domain wall profile and separately treat three qualitatively different dynamical regimes.
We conclude by comparing our findings with the analogous problem in the (integrable) quantum Heisenberg (anti)ferromagnet, and 
highlight a remarkable classical-quantum correspondence for the macroscopic spin transport.



\paragraph*{Landau--Lifshitz model.}
The Landau--Lifshitz model is a classical field theory which governs a precessional motion of spin field on the unit sphere,
described by the equation of motion~\cite{Lakshmanan76,Lakshmanan77,Takhtajan77,Sklyanin79,Fogedby80,Faddeev_book}
\begin{equation}
\vec{S}_{t} = \vec{S}\times \vec{S}_{xx} + \vec{S}\times {\bf J}\,\vec{S},\qquad \vec{S}\cdot \vec{S} = 1,
\label{eqn:EOM}
\end{equation}
with $\vec{S}\equiv (S^{x},S^{y},S^{z})^{\rm T}$. Choosing the uniaxial anisotropy tensor
${\bf J}\equiv {\rm diag}(0,0,\delta)$, there are three regions to be distinguished by the value of the  parameter
$\varepsilon~\equiv~\ii\sqrt{\delta}$: the easy-axis regime $ \varepsilon^2 < 0$, the easy-plane
regime $\varepsilon^2 > 0$, and the isotropic case $\varepsilon = 0$. 
This model also appears in a long-wavelength description of the spinor Bose gases \cite{PhysRevA.77.063622,PhysRevA.82.053614}.

\paragraph*{Spin transport.}
To study spin transport, we consider the initial profile in the form of a (smooth) domain wall of the width $x_{0}$,
\begin{equation}
\vec{S}(x,t=0) = \big({\rm sech}{(x/x_{0})},0,\tanh{(x/x_{0})}\big)^{\rm T},
\label{eqn:domain_wall}
\end{equation}
which connects two distinct (degenerate) vacua. With no loss of generality we can put $x_{0} = 1$ by a simple rescaling
$x\to x_{0}\,x$, $t\to x_{0}^{2}t$, $\varepsilon\to \varepsilon/x_{0}$.

To characterize spin dynamics, it is natural to use a dynamical quantity
\begin{equation}
m(t) = \int^{\infty}_{0}\dd x\,\big(1-S^{z}(x,t)\big),
\label{eqn:m}
\end{equation}
which measures the change of total magnetization in the right half-system and
has been already employed in previous studies \cite{Ljubotina17,Misguich17}.

Eq.~\eqref{eqn:EOM} is completely integrable and thus possesses infinitely many conserved charges.
Spin density $S^{z}$ corresponds to the globally conserved Noether charge and should be distinguished from
other charges (the momentum, energy, and higher charges) which are all initially localized at the domain boundary and undergo ballistic spreading, in exact analogy to the expansion of local conserved charges in the nonlinear Schr\"{o}dinger equation \cite{MIG17}.

\paragraph*{Nonlinear Fourier transform.}

The standard procedure to integrate nonlinear integrable wave equations such as Eq.~\eqref{eqn:EOM} is called
the inverse scattering method. We briefly sketch the main relevant ideas below, while for the full description we refer to
one of the standard textbooks~\cite{Faddeev_book,Ablowitz_book,Novikov_book}.

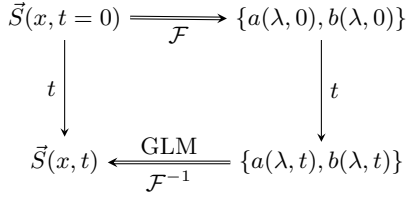
\begin{figure}[htb]
\centering
\begin{tikzpicture}
  \matrix (m) [matrix of math nodes,row sep=4em,column sep=4em,minimum width=3em]
  {
     \vec{S}(x,t=0) & \{a(\lambda,0),b(\lambda,0)\} \\
     \vec{S}(x,t) & \{a(\lambda,t),b(\lambda,t)\} \\};
  \path[-stealth]
    (m-1-1) edge node [left] {$t$} (m-2-1)
            edge [double] node [below] {$\mathcal{F}$} (m-1-2)
    (m-1-2) edge node [right] {$t$} (m-2-2);
  \path[stealth-]  
  	(m-2-1.east|-m-2-2) edge [double] node [below] {$\mathcal{F}^{-1}$} node [above] {GLM} (m-2-2);
\end{tikzpicture}
\caption{Schematic representation of the integration protocol: the forward `nonlinear Fourier transform' $\mathcal{F}$,
maps the initial spin field $\vec{S}(x,0)$ to the spectral data $\{a(\lambda),b(\lambda)\}$. The latter
satisfies simple time-evolution \eqref{eqn:time_evolution}. The inverse transform $\mathcal{F}^{-1}$
amounts to solving an appropriate linear integral equation from where one reconstructs the time-evolved spin-field $\vec{S}(x,t)$.}
\label{fig:scheme}
\end{figure}

The framework of integrability relies on a geometric picture of linear parallel transport for the auxiliary
wavefunction $\psi=\psi(x,t)$,
\begin{equation}
\begin{split}
\partial_{\sigma}\psi(\lambda;x,t) = {\bf U}_{[\sigma]}(\lambda;x,t)\psi(\lambda;x,t),
\end{split}
\label{eqn:linear_problem}
\end{equation}
for $\sigma \in \{x,t\}$, and the spatial and temporal connection components are
\begin{align}
{\bf U}_{[x]}(\lambda) &= \frac{1}{2\ii}\sum_{\alpha}w_{\alpha}S^{\alpha}\boldsymbol{\sigma}^{\alpha},\\
{\bf U}_{[t]}(\lambda) &= \frac{1}{2\ii}\sum_{\alpha}\!\left[w_{\alpha}(\vec{S}\times \vec{S}_{x})^{\alpha} \boldsymbol{\sigma}^{\alpha}
\!-\!\frac{w_{x}w_{y}w_{z}}{w_{\alpha}}S^{\alpha}\boldsymbol{\sigma}^{\alpha}\right],
\end{align}
respectively~\cite{Sklyanin79,Faddeev_book,BBI14}.
Here, $w_{x}=w_{y}=\sqrt{\lambda^2-\varepsilon^{2}}$, $w_{z}=\lambda$, and $\lambda$ is the spectral parameter on a two-sheeted
Riemann surface $\mu(\lambda)=\sqrt{\lambda^2-\varepsilon^{2}}$. Eq. \eqref{eqn:EOM} follows from the zero-curvature condition
$[\partial_{x}-{\bf U}_{[x]},\partial_{t}-{\bf U}_{[t]}]=0$, which is needed for the consistency of
Eqs.~\eqref{eqn:linear_problem}. Imposing the initial condition \eqref{eqn:domain_wall}, we construct two Jost solutions of the 
spatial part of Eqs.~\eqref{eqn:linear_problem} ${\bf T}_{\pm}$, characterized by asymptotic behavior
${\bf T}_{+}(x\to \infty)=\exp{(\lambda x \boldsymbol{\sigma}^{z}/2\ii)}$ and
${\bf T}_{-}(x\to -\infty)=\exp{(-\lambda x \boldsymbol{\sigma}^{z}/2\ii)}\ii \boldsymbol{\sigma}^{x}$. The transfer matrix 
${\bf T}(\lambda;t)$ is defined as a unimodular 
constant matrix that interpolates between Jost solutions ${\bf T}_{-}={\bf T}_{+}{\bf T}(\lambda)$. It can be presented as
\begin{equation}
{\bf T}(\lambda) = \begin{pmatrix}
a(\lambda) & -\bar{b}(\lambda) \\
b(\lambda) & \bar{a}(\lambda)
\end{pmatrix}.
\end{equation}
Complex functions $a(\lambda)$ and $b(\lambda)$ are called scattering amplitudes and store full information about the initial profile. 
The scattering data satisfy simple time-evolution
\begin{equation}
a(\lambda,t)=a(\lambda,0),\quad b(\lambda,t)=b(\lambda,0)e^{\ii(\lambda^{2}-\varepsilon^{2})t},
\label{eqn:time_evolution}
\end{equation}
which can be inferred from the temporal part of Eqs.~\eqref{eqn:linear_problem}. The conserved charges can be expressed
as moments of the `density of states' $\rho(\lambda)=\log|a(\lambda)|^2$.

The solution to Eqs.~\eqref{eqn:linear_problem} for the domain wall profile \eqref{eqn:domain_wall} leads to the following scattering data
\begin{align}
a(\lambda,0) &= \frac{\sqrt{\lambda^{2}-\varepsilon^{2}}\,\Gamma^{2}(\tfrac{1}{2}-\tfrac{\ii}{2}\lambda)}
{2\Gamma(1-\tfrac{\ii}{2}(\lambda-\varepsilon))\Gamma(1-\tfrac{\ii}{2}(\lambda+\varepsilon))},\\
b(\lambda,0) &= \ii \frac{\cosh{(\tfrac{\pi}{2}\varepsilon)}}{\cosh{(\tfrac{\pi}{2}\lambda)}}.
\end{align}
The time-evolution for the spin field can be restored from the scattering data \eqref{eqn:time_evolution} by
the inverse transform, shown in Fig.~\ref{fig:scheme}. The latter takes the form of a linear integral Fredholm-type equation called the
Gel'fand--Levitan--Marchenko (GLM) equation~\cite{Faddeev_book}. Its precise form depends crucially on the value $\varepsilon$
and the type of boundary condition adjoined to Eq.~\eqref{eqn:EOM}. The presented analysis is confined to the non-trivial 
topological sector of the theory which requires certain (sometimes subtle) adaptations of the standard procedure \cite{GMI_long}.

\begin{figure*}[htb]
\centering
\includegraphics[width = 0.9\textwidth]{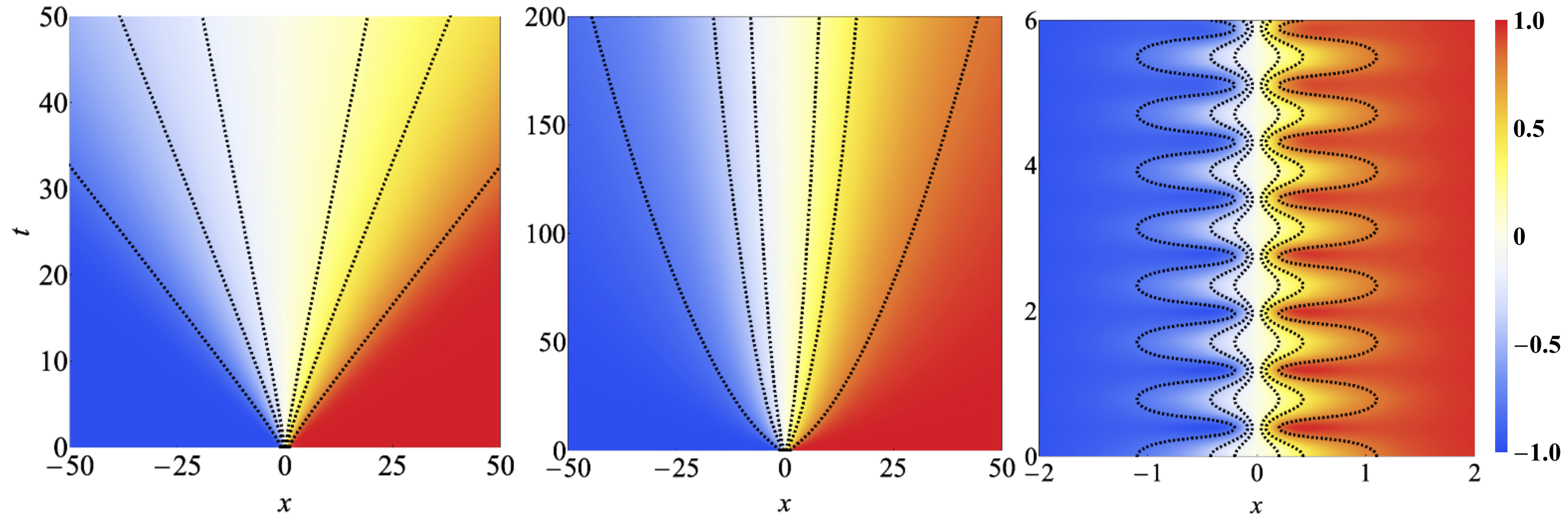}
\caption{Time-dependent density profiles of $S^z$ component in the easy-plane $\delta=- 1$ (left), isotropic $\delta=0$ (middle)
and easy-axis $\delta=9$ (right) regimes, displaying ballistic spin transport, logarithmically enhanced diffusion and absence of transport, 
respectively. The dashed lines show $|S^{z}|~=~\{0.2, 0.4, 0.8\}$.}
\label{fig:Density_Plot}
\end{figure*}

\paragraph*{Easy-plane regime.}

The absence of zeros of $a(\lambda)$ in the upper-half $\lambda$-plane for $\varepsilon \in \mathbb{R}$ means
that the spectrum comprises only from a dispersive continuum of radiative modes. In fact, the origin of ballistic transport can
be explained without recourse to the exact solution. It suffices to consider a hydrodynamic approximation to the
equation of motion \cite{Whitham_book}. Introducing slow variables $S^{z}$ and $v \equiv -\ii[\log S^{+}(x)]_{x}$,
along with the nonlinearity $R \equiv [(S^{z}_{x}/\sqrt{1-(S^{z})^{2}})_{x}(1/\sqrt{1-(S^{z})^{2}})]_{x}$,
Eq.~\eqref{eqn:EOM} can be put in the form
\begin{equation}
S^{z}_{t} - [(1-(S^{z})^{2})v]_{x} = 0,\quad
v_{t} - [(\varepsilon^{2}-v^{2})S^{z}]_{x} = R.
\end{equation}
By disregarding the nonlinearity term $R$, one has
\begin{equation}
\begin{pmatrix}
S^{z} \\ v
\end{pmatrix}_{t} =
\begin{pmatrix}
-2S^{z}v & 1-(S^{z})^{2} \\
\varepsilon^{2}-v^{2} & -2S^{z}v
\end{pmatrix}
\begin{pmatrix}
S^{z} \\ v
\end{pmatrix}_{x}.
\label{eqn:hydrodynamic_system}
\end{equation}
This WKB-type approximation can alternatively be viewed as the simplest case of a more general Whitham theory describing modulation
of multiphase solutions to nonlinear wave equations~\cite{Whitham_book}. The system \eqref{eqn:hydrodynamic_system} can be brought into the Riemann diagonal form $\partial_{t}r_{\pm}(x,t)+V_\pm(x,t)\partial_{x}r_{\pm}(x,t)=0$, with Riemann invariants
$r_{\pm}=S^{z}v\pm \sqrt{(1-(S^{z})^{2})(\varepsilon^{2}-v^{2})}$, and characteristic velocities $V_{+}=r_{-}/2+3r_{+}/2$, $V_{-}=3r_{-}/2+r_{+}/2$.

The absence of scale in the initial profile motivates one to seek for the self-similar solution depending on the ray coordinate
$\xi = x/t$, which yields the hydrodynamic equation
$[V_{\pm}(\xi)-\xi]\partial_{\xi}r_{\pm}(\xi) = 0$.
To single out a unique solution, we need to additionally supply appropriate boundary conditions, which are set by the values 
$S^{z}(\xi_{\pm})=\pm 1$ and $v(\xi_{\pm})=v_{0}={\rm const}$ at $\xi_{\pm}$ -- boundaries  of the ballistically
expanding region connecting two vacua. Inside this region the solution reads
\begin{equation}
S^{z}(\xi) = \frac{\xi}{2|\varepsilon|},\qquad v = |\varepsilon| = v_0,\qquad \xi_{\pm}=\pm 2|\varepsilon|,
\end{equation}
which implies linear growth of magnetization \eqref{eqn:m}, namely
$m(t) \simeq t \int^{2|\varepsilon|}_{0}\dd \xi\,(1-S_z(\xi)) = |\varepsilon|t$.
Notice that the density of states $\rho(\lambda)$ develops a singularity at $\lambda_{*}=|\varepsilon|$, which
thus defines a natural scale in the spectrum. The velocity of the hydrodynamic region is nothing but
the velocity of the critical dispersive modes $v_{*}=2\lambda_{*}=|\xi_{\pm}|$.
Moreover, a non-trivial solution on Euler scale exists only strictly in the easy-plane regime $\varepsilon^{2}>0$, whereas for 
$\varepsilon^{2}\leq 0$ the hydrodynamic solution trivializes, implying sub-ballistic transport.

\paragraph*{Isotropic interaction.}

For $\varepsilon=0$, the density of states $\rho(\lambda)$ logarithmically diverges at $\lambda\to 0$. As we demonstrate,
this turns out to be an  artefact of the specific domain-wall profile with perfectly anti-parallel asymptotic spin fields.
For this reason, we also consider a deformed profile
$\vec{S}=(\cos{\Phi},0,\sin{\Phi})^{\rm T}$, where $\Phi=(\gamma/\pi)\arcsin{(\tanh{x})}$ with
the `twisting angle' $\gamma\in [0,\pi)$. The induced correction to the 
scattering data for $\gamma\approx \pi$, computed with the first order perturbation theory, displaces
the zero of $a(\lambda)$ at the origin, $a(0) \approx \ii (\pi -\gamma)/2$, rendering the density of states finite.

\begin{figure}[b]
\centering
\includegraphics[width=0.47\textwidth]{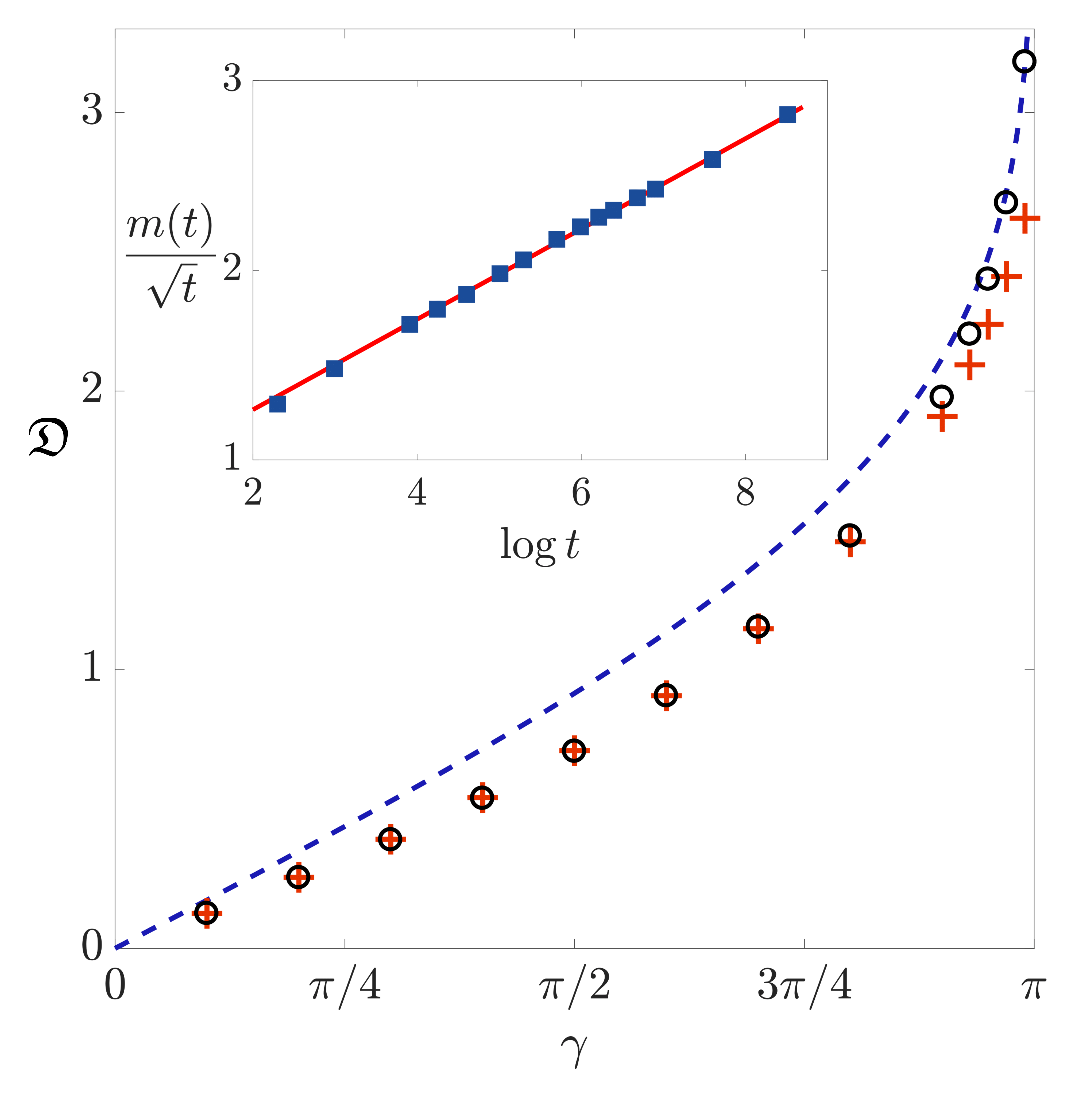}
\caption{Spin diffusion constant $\DD(\gamma)$ as a function of the
twisting angle $\gamma$, shown for the self-similarity solutions (open circles) and numerical integration up to $t=2000$~(red crosses). 
Blue dashed line shows the leading term in the large-$E$ asymptotic expansion of the self-similar solution.
Inset: Numerical solution to the inverse scattering transform of the untwisted domain wall profile \eqref{eqn:domain_wall}.}
\label{fig:figure3}
\end{figure}

At the isotropic point, there is a unique class of self-similar solutions to Eq.~\eqref{eqn:EOM} which depend
on the scaling variable $\zeta=x/\sqrt{t}$, governed by an ODE~\cite{Lakshmanan76},
\begin{equation}
-2\zeta\vec{S}_{\zeta} = \vec{S}\times \vec{S}_{\zeta\zeta},
\label{eqn:self-similar}
\end{equation}
which is usually studied in the context of the vortex filament dynamics~\cite{GRV03}.
For initial conditions with a jump discontinuity at the origin, Eq. \eqref{eqn:self-similar} can be solved analytically.
For large times, we observe that the twisted domain wall approaches the self-similar profile.
The latter manifestly yields normal spin diffusion $m(t)\sim \DD(\gamma) \sqrt{t}$. The diffusion constant~\footnote{This
definition of the diffusions constant should not be confused with that of the Kubo linear response theory.} 
$\DD(\gamma)$ plays a role of the filament curvature and can be approximated as $c\sqrt{E}$, with
$c=\sqrt{2}(\pi -2\log(\sqrt{2}+1))\approx 2$  and $E= \vec{S}_{\zeta}^2$ 
being the conserved energy \cite{GL2019}. Using relation $e^{-\pi E/2}=\cos(\gamma/2)$, one concludes that $\DD(\gamma)$ diverges as $\gamma\to \pi$,
explaining the breakdown of normal diffusion for the untwisted profile \eqref{eqn:domain_wall}.
In order to quantify it, we have implemented an efficient numerical solver of the inverse (GLM) transform 
$\mathcal{F}^{-1}$ (see Fig.~\ref{fig:scheme}). Our data indicates a mild logarithmic (in time) divergence of $m(t)$
(see Fig.~\ref{fig:figure3}, inset plot), which nicely conforms with the type of singularity in the density of states. The twist of 
the boundary conditions removes the singularity and restores normal spin diffusion, as shown in Fig.~\ref{fig:figure3}.

\paragraph*{Easy-axis regime.}
In distinction to the previous two regimes, the scattering data acquire an additional discrete component which physically corresponds 
to the  (multi)soliton modes. The simplest among them are static (anti)-kink modes with topological charge $Q=\pm 1$, which 
coincide with domain wall \eqref{eqn:domain_wall} for $x_{0} = \pm 1/\sqrt{\delta}$. The kink persists in the spectrum for all
$\delta > 0$. Besides solitons, the spectrum involves a continuous spectrum of radiative modes, which, however, vanish for the the 
discrete set of `reflectionless anisotropies' $\varepsilon = \ii(2n+1)$,
$n \in \mathds{Z}$. The analyticity of $a(\lambda)$ can be restored with the uniformization map,
$\lambda(z)=(z+\varepsilon^{2}z^{-1})/2$; soliton modes are then characterized by zeros of $a(z)$ located in the upper half $z$-plane.
The spectrum of the domain wall does not involve any asymptotically free solitons, implying a trivial ballistic channel.
The asymptotic scaling $m(t) \sim t^0$ is then a consequence of the finite difference between the domain wall profile and
the stable kink. For instance, on the interval $0<\varepsilon < 3\ii$, the kink is the only soliton mode and thus the steady state
of the domain wall dynamics. On the other hand, for larger values of anisotropy we obtained an infinite family of bound states which 
undergo periodic oscillatory motion.
To our knowledge, such solutions have not been explicitly described previously in the
literature~\cite{Kosevich_review,BogdanKovalev,Svendsen93,BBI14}, but similar `wobbling kinks' have been already identified 
in the sine--Gordon model~\cite{Segur83,DeRossi98,Kalbermann04,Ferreira08}.
For example, for $n=1$ the scattering data read
\begin{equation}\label{wobble}
a(z)=\ii \frac{(z-3\ii)(z^2-2\ii z+9)}{(z+3\ii)(z^2+2\ii z+9)},\quad b(z)=0,
\end{equation}
and describes the kink-breather bound state which can be compactly parametrized by a complex stereographic angle $\varphi$,
\begin{equation}
S^{z} = \frac{1-|\varphi|^{2}}{1+|\varphi|^{2}},\qquad S^{x} + \ii S^{y} = \frac{2\varphi}{1+|\varphi|^{2}},
\label{eqn:reconstruction}
\end{equation}
reading
\begin{equation}
\varphi = \frac{e^{\eta_{0}}+e^{\eta_{+}}+2e^{\eta_{-}}}{1 + 2e^{\eta_{0}+\eta_{-}}+e^{\eta_{0}+\eta_{+}}}.
\label{eqn:kink-breather}
\end{equation}
The phases $\eta_{i}(x,t)=\ii(k_{i}x+\omega_{i}t)$ and $k_{0}=-3\ii$, $\omega_{0}=0$, and $k_{\pm}=\pm \ii$ and
$\omega_{\pm}=k^{2}_{\pm}-\varepsilon^{2}$ are determined from the scattering data \eqref{wobble}. 
The full classification of the soliton spectrum is postponed to \cite{GMI_long}.

\paragraph*{Classical-quantum correspondence.}

The quantum integrable (lattice) counterpart to the equation of motion \eqref{eqn:EOM} is the celebrated anisotropic quantum
Heisenberg spin chain
$H~\simeq~-\sum_{i}\left(\hat{S}^{x}_{i}\hat{S}^{x}_{i+1}+\hat{S}^{y}_{i}\hat{S}^{y}_{i+1}
+\Delta \hat{S}^{z}_{i}\hat{S}^{z}_{i+1}\right)$,
the oldest known model solvable by the Bethe Ansatz~\cite{Bethe31,Takahashi71,Gaudin71,TS72}.
The time-evolution following a sharp magnetic domain and its dependence on anisotropy $\Delta$
has already been a subject of study in the past~\cite{Gobert05,Mossel_2010,BCDF16,Collura_DW,Ljubotina17,Misguich17,Ljubotina_JPA}.

In the remainder of the Letter, we wish to elaborate on the perfect \emph{qualitative} agreement in the spin dynamics of
the classical and quantum anisotropic ferromagnets, in spite of rather discernible differences in the respective microscopic dynamics:
the spectrum of excitations of quantum dynamics (classified in \cite{Takahashi71,TS72}) consists of magnons (and bound states thereof) carrying a quantized amount of spin, whereas classical dynamics corresponds to the semi-classical long-wavelength spectrum of large spin-coherent states~\cite{Sutherland95,DS00,KMMZ04,Bargheer08}.

To facilitate the comparisons, we briefly review the key known results.
Ballistic expansion of the magnetic domain wall in the gapless regime $|\Delta|<1$ has been first computed numerically using the 
hydrodynamic theory for quantum integrable models \cite{BCDF16} and latter obtained analytically in \cite{Collura_DW}.
The dynamical freezing of the magnetic domain wall in the gapped regime $|\Delta|>1$ has been reported
in \cite{Gobert05,Ljubotina17,Misguich17}. In fact, the observed effect is once again a consequence of 
stable topological kink vacua, representing an inhomogeneous (infinite-volume) ground states with a finite spectral
gap \cite{Alcaraz95,Koma97} (which become unstable at $\Delta = 1$).
At the isotropic point, the observed logarithmically enhanced diffusion law in the isotropic Landau--Lifshitz model (cf. Fig.~\ref{fig:figure3}) 
appears to be compatible with the state-of-the-art numerical study \cite{Misguich17} (which is missed, somehow, 
in \cite{Ljubotina_JPA}). Curiously, the same type of correction has been found in the asymptotic behavior of the return probability 
amplitude for the domain wall initial state \cite{Stephan17}.
Our twisted domain wall profile should be understood as a classical analogue of the tilted domain wall product 
states employed in \cite{Ljubotina_JPA} which exhibit normal spin diffusion.

Although in this Letter we concentrated solely on the spin dynamics in the far-from-equilibrium regime (with a specific initial 
state), there exists a robust evidence that the classical-quantum correspondence holds also in thermal equilibrium in the conventional 
framework of linear response theory. The thermal spin diffusion constant (at half filling) 
in the lattice Landau--Lifshitz model -- defined via the thermal average of the time-dependent auto-correlation
$C(t)=\expect{J(0)J(t)}/L$ of the spin current $J(t)$ -- has been numerically investigated in \cite{PZ13},
where three distinct regimes have been identified: ballistic transport with a finite Drude weight 
$\mathcal{D}=\lim_{t\to \infty}C(t)$ in the easy-plane regime, normal diffusion with finite
$D=\lim_{t\to \infty}\int^{t}_{0}C(t^{\prime})\dd t^{\prime}$ in the easy-axis regime, and
superdiffusion with a time-dependent diffusion constant $D(t)\sim t^{1/3}$ at the isotropic point.
In the quantum Heisenberg spin-$1/2$ chain the picture remains qualitatively the same: in the easy-plane regime ($|\Delta|<1$),
the finite spin Drude weight has been attributed to hidden quasi-local conservation laws~\cite{PI13,Ilievski_review} and computed 
exactly in \cite{IDN_Drude,Hubbard_Drude} using the hydrodynamic theory for integrable models~\cite{CDY16,BCDF16}.
In the easy-axis regime ($|\Delta|>1$) one finds normal diffusion, theoretically explained
in \cite{DeNardis_diffusion,DeNardis_diffusion_long}.
Finally, the divergence of the spin diffusion constant at the isotropic point (at finite temperature and half filling) has been
established in \cite{superdiffusion}. Numerical simulations \cite{Ljubotina17} provide a convincing evidence for super-diffusion
with the Kardar-Parisi-Zhang~(KPZ) dynamical exponent $\alpha = 2/3$, later theoretically justified with the aid of a dimensional analysis in \cite{GV18}.

\paragraph*{Conclusions.}

We have studied the spin transport in the uniaxial Landau--Lifshitz ferromagnet initialized in the domain wall profile.
We have computed the exact spectrum of nonlinear normal modes and expressed the
time-evolved spin field as a solution of the inverse scattering transformation.

In the easy-plane regime we encountered ballistic expansion, which, to the leading order,
can be captured by a simple hydrodynamic theory.

For the isotropic interaction, we rigorously established a divergent spin diffusion constant and explain the origin of
the a modified diffusion law with a multiplicative logarithmic correction. The effect is shown to be a particularity of the 
initial state and can be regularized by a twist of the boundary conditions which restores normal diffusion. Such a `$\pi$-anomaly'
can be understood as an `infrared catastrophe' due to a logarithmic divergence of the mode occupation function in the
low-energy $\lambda \to 0$ limit.

In the easy-axis regime, the spectrum of the domain wall acquires non-trivial topologically charged (multi)soliton states
which consists of breather modes superimposed on a kink. It remains an interesting open question whether wobbling kinks survive
quantization, similar to the problem of quantum stability of cnoidal waves addressed in \cite{Bargheer08}.
Analytic continuation into the easy-plane phase, $\varepsilon \to -\ii \varepsilon$, can also be understood as destabilization of the kink mode into a dynamical domain wall.

Since the Landau--Lifshitz model can be regarded as a generic integrable $(1+1)$-dimensional soliton system, it is compelling to conjecture that the correspondence is a general feature of quantum integrable lattice models that admit a semi-classical limit 
(such as e.g. the sine--Gordon model, nonlinear sigma models, nonlinear Schr\"{o}dinger equation, etc.). We hope that our results 
can stimulate further research in this direction. A separate interesting issue is whether similar correspondences
appear even more broadly, e.g. in one of the non-integrable dynamical systems in higher dimensions.

\textit{Acknowledgments.}
We thank O. Lisovyy, J. De Nardis, M. Medenjak and T. Prosen for their comments.
Y. M. and O. G. acknowledge the support from the European Research Council under ERC Advanced grant 743032 DYNAMINT. E. I. is supported by VENI grant number 680-47-454 by the Netherlands Organisation for Scientific Research (NWO).
\bibliography{LLDW}
\bibliographystyle{apsrev4-1}

\end{document}